\documentclass[aps,pre,twocolumn,amsmath,amssymb,floatfix,footinbib,superscriptaddress,letterpaper]{revtex4-1}

\usepackage{amsmath,amssymb,mathtools}
\usepackage{stmaryrd}
\usepackage{times}
\usepackage[euler]{textgreek}
\usepackage[bbgreekl]{mathbbol}
\usepackage{graphicx}
\usepackage{bm,color}
\usepackage{natbib,hyperref}
\usepackage[capitalize]{cleveref}

\hypersetup{colorlinks=true,linkcolor=blue,citecolor=blue,urlcolor=blue}

\begin{document}

\title{Breaking optomechanical cooling limit by two drive fields on a membrane-in-middle system}
\author{Chuan Wang}
\affiliation{Fujian Key Laboratory of Light Propagation and Transformation, College of Information Science and Engineering, Huaqiao University,
Xiamen 361021, China}
\author{Qing Lin}
\email{qlin@hqu.edu.cn}
\affiliation{Fujian Key Laboratory of Light Propagation and Transformation, College of Information Science and Engineering, Huaqiao University,
Xiamen 361021, China}
\author{Bing He}
\email{bing.he@umayor.cl}
\affiliation{Center for Quantum Optics and Quantum Information,
Universidad Mayor, Camino La Pir\'{a}mide 5750, Huechuraba, Chile}



\begin{abstract}
We present a theoretical scheme for ground state cooling of a mechanical resonator in a membrane-in-middle optomechanical system (OMS) driven by two red-detuned drive fields. The details of dynamical evolution of OMS are provided, and the effect of system conditions on cooling results are systematically studied. Most importantly, the setups with two drives are found to be capable of achieving better cooling results than the theoretical cooling limit with single cavity. Even an improvement by one order of thermal phonon number is possible with proper combination of the cavity damping rate and drive intensity.
\end{abstract}

\maketitle
\section{Introduction}

Ground state cooling is a method to create macroscopic quantum state, which can test the fundamental problem of whether there exists a boundary between the classical and quantum worlds \cite{decoherence,boundary1,boundary2}. Recently, optomechanical system (OMS) has been used to realize ground state cooling of a mechanical resonator under the radiation pressure of cavity modes \cite{OMS1}. Various developments in theories (see, e.g. \cite{coolt1,coolt2,coolt3,coolt4,coolt5,coolt6,coolt7,coolt8,coolt9,coolt10,coolt11,coolt12,
coolt13, cooling}) and experiments (see, e.g. \cite{coole1,coole2,coole3,coole4,coole5,coole6,coole7,coole8,coole9,coole10,coole11,coole12,coole13,coole14}) have been reported over the past few years.

To any realistic OMS, however, the mechanical resonator can not be cooled down to arbitrary extent, because of its interaction with thermal environment and the inherent noise effects. For example, the experimentally achieved thermal phonon number of a mechanical resonator was found to be incapable of being lowered further with increased cooling laser power \cite{coole8}. In the presence of a thermal environment, the cooling limit of an OMS under the drive of red-detuned laser was predicted to be $n_{m,f}=(\gamma_m/\Gamma_{eff})n_{th}$, a fraction of the environmental thermal phonon number $n_{th}$ by the ratio of the mechanical damping rate $\gamma_m$ and an effective cavity damping rate $\Gamma_{eff}$, in addition to an inherent remnant due to the action of quantum noises \cite{OMS1}. 
It was more recently proved with the dynamical evolution of OMS that the cooling limit is exactly $n_{m,f}=(\gamma_m/\kappa)n_{th}$ ($\kappa$ is the cavity damping rate) for arbitrary OMS under a single red-detuned cooling laser \cite{cooling}. 
For an OMS under single cooling drive, therefore, the only possible way to achieve better cooling result is to have a high quality factor of the mechanical resonator, which relies on the relevant technology advances. Whether such restriction can be avoided or 
relaxed in other ways is significantly meaningful 
to optomechanical cooling.

In the current work we consider a kind of modified OMS, i.e. membrane-in-middle (MIM) systems (see Fig. 1), for the improvement of cooling. 
So far some physical properties of such systems have been investigated \cite{mim1,mim2,mim3}. In 2011, Li et al. suggested to realize ground state cooling with one red-detuned and one blue-detuned continuous-wave (CW) drive fields in the MIM system \cite{mim4}. In the following year, a theoretical and experimental investigation of cooling in the MIM system under a single CW drive field and the cooling 
to the effect of $n_{m,f}\sim 14 (\gamma_m/\kappa) n_{th}$ were reported \cite{mim5}. Some other related theoretical schemes \cite{mim6, mim7, mim8} were also proposed.

However, no systematic investigation of cooling with the MIM systems, regarding how the systems' parameters affect their cooling performance, was reported in the past. In particular, the detailed cooling process with a MIM system had not been discussed; hence, the question whether or not it is better than the one with only one single cavity was still open. Here we address the problem by presenting the details of the cooling processes with MIM system under two red-detuned CW drives. Moreover, we provide the relations between the cooling effect and system conditions to demonstrate that the theoretical cooling limit $n_{m,f}=(\gamma_m/\kappa)n_{th}$ with single cavity can be surpassed (even one order improvement, i.e., $n_{m,f}=0.1(\gamma_m/\kappa)n_{th}$) with a MIM system, which has a proper combination of cavity damping rate and drive intensity in the second cavity. 

The rest of the paper is organized as follows. In Sec. \ref{sec2}, the MIM OMS Hamiltonian is linearized with the approach used in Ref. \cite{cooling} and the validity of that approach is discussed as well. With the derived dynamical equations, cooling with two red-detuned drives in MIM system is presented in Sec. \ref{sec3}. To clearly understand how the system conditions affect the cooling effect, we discuss the relations between the cooling ratio and the systems parameters, including the cavity damping rate, tunneling rate, and effective drive intensity in Sec. \ref{sec4}. Furthermore, in Sec. \ref{sec5}, the analytical solutions of cooling limit under different conditions are provided. Finally this work is concluded in Sec. \ref{conclusion}.

\section{System Hamiltonian and dynamical equations}
\label{sec2}

\begin{figure}[b]
\centering\includegraphics[width=9cm]{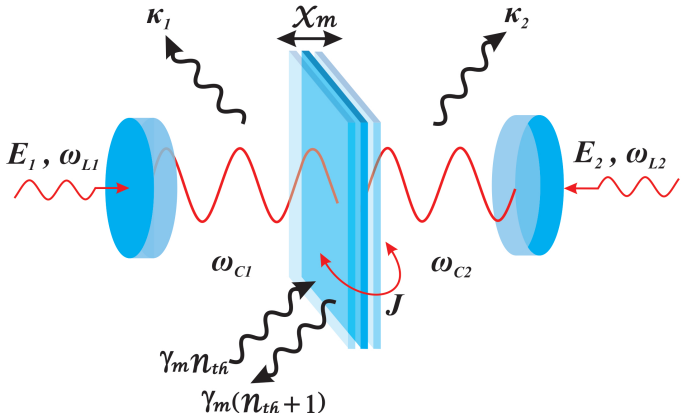}
\caption{Model of optomechanical cooling with the membrane-in-middle system. The two continuous wave drive fields with red detuning $\Delta_1=\Delta_2=\omega_m$ are used to cool the same mechanical resonator (the membrane inside the cavity). A more efficient cooling effect than that with only one cavity can be obtained by choosing proper conditions.}  \label{MIM}
\end{figure}

We consider an OMS, shown in Fig. \ref{MIM}, with a thin membrane (the mechanical resonator with resonant frequency $\omega_m$) placed inside a normal cavity. Two CW laser beams with the same central frequency $\omega_{L}$ and constant amplitude $E_1$ and $E_2$ from the left and right cavities, respectively, drive the membrane. The Hamiltonian of the system reads ($\hbar=1$),
\begin{align}
H(t)=H_S(t)+H_{OM}+H_{SR}(t)
\end{align}
where
\begin{align}
H_S(t)=&\omega_{c_1} \hat{a}_1^{\dag}\hat{a}_1+\omega_{c_2} \hat{a}_2^{\dag}\hat{a}_2+\omega_m  \hat{b}^{\dag}\hat{b}\nonumber\\
&+i[\hat{a}_1^{\dag}E_1e^{-i\omega_L t}-\hat{a}_1E_1^*e^{i\omega_L t}]\nonumber\\
&+i[\hat{a}_2^{\dag}E_2e^{-i\omega_L t}-\hat{a}_2E_2^*e^{i\omega_L t}],\\
H_{OM}=&-g_{m}(\hat{a}_1^{\dag}\hat{a}_1-\hat{a}_2^{\dag}\hat{a}_2)(\hat{b}+\hat{b}^{\dag})+J(\hat{a}_2^{\dag}\hat{a}_1+\hat{a}_1^{\dag}\hat{a}_2), \\
H_{SR}(t)=&i\sqrt{2\kappa_1}\{\hat{a}_1^{\dag} \hat{\xi}_{c_1}(t)-\hat{a}_1\hat{\xi}^{\dag}_{c_1}(t)\}\nonumber\\
&+i\sqrt{2\kappa_2}\{\hat{a}_2^{\dag} \hat{\xi}_{c_2}(t)-\hat{a}_2\hat{\xi}^{\dag}_{c_2}(t)\}\nonumber\\
&+i\sqrt{2\gamma_m}\{\hat{b}^{\dag} \hat{\xi}_m(t)-\hat{b}\hat{\xi}^{\dag}_m(t)\}.
\end{align}
The first part $H_S(t)$ involves free Hamiltonians of the mechanical mode and two cavity modes having resonant frequency $\omega_{c_1}$ and $\omega_{c_2}$, respectively, with the two external drives. The second part $H_{OM}$ is the coupling of the two cavity modes with the mechanical mode due to the radiation pressure of single-photon coupling strength $g_m$, and the coupling of the two cavity modes themselves, due to the tunneling, with the coupling strength $J$. The cavity length of the left and right cavities are changed in opposite manner, resulting in the opposite signs of coupling terms in $H_{OM}$.
The final part $H_{SR}$ describes the couplings of the cavity and mechanical modes with the environmental reservoirs. The corresponding stochastic Langevin noise operator $\hat{\xi}_c$ ($\hat{\xi}_m$) of the reservoir satisfies the relation $<\hat{\xi}^\dagger_l(t)\hat{\xi}_l(\tau)>_R=n_l\delta(t-\tau)$ ($l=c,m$) with the occupation number $n_l$ in thermal equilibrium condition. Here, we assume that $n_c=1/(e^{\hbar \omega_c/k_B T}-1)\approx 0$ and $n_m=1/(e^{\hbar \omega_m/k_B T}-1)=n_{th} \gg 1$, where $n_{th}$ is the thermal phonon number.

The evolution of such OMS can be described by the generalized evolution operator $U(t)=\mathcal{T}\exp\{-i\int_0^t d\tau \hat{H}(\tau)\}$ as a time-ordered exponential \cite{book}, leading to the following nonlinear quantum Langevin equations: 
\begin{align}
\dot{\hat{a}}_1=&-\kappa_1 \hat{a}_1+i g_m (e^{-i\omega_mt}\hat{b}+e^{i\omega_mt}\hat{b}^{\dag})\hat{a}_1\nonumber\\
&-iJ\hat{a}_2+E_1e^{i\Delta_1 t}+\sqrt{2\kappa_1}\hat{\xi}_{c_1} (t),\nonumber\\
\dot{\hat{b}}=&-\gamma_m \hat{b}+i g_m e^{i\omega_m t}(\hat{a}_1^{\dag}\hat{a}_1-\hat{a}_2^{\dag}\hat{a}_2)+\sqrt{2\gamma_m}\hat{\xi}_m(t),\nonumber\\
\dot{\hat{a}}_2=&-\kappa_2 \hat{a}_2-i g_m (e^{-i\omega_mt}\hat{b}+e^{i\omega_mt}\hat{b}^{\dag})\hat{a}_2\nonumber\\
&-iJ\hat{a}_1+E_2e^{i\Delta_2 t}+\sqrt{2\kappa_2}\hat{\xi}_{c_2} (t),\label{neq}
\end{align}
where the detunings are $\Delta_{1(2)}=\omega_{c_{1(2)}}-\omega_L$. The nonlinear terms in each quantum Langevin equation make it challenging to arrive at the analytical solution. The common approach is to decompose the cavity field modes $\hat{a}_{1(2)}$ to the sum of the mean value (steady state) $\alpha_{1(2)}(t)$ and their quantum fluctuations $\delta \hat{a}_{1(2)}$, and then linearize the above equations \cite{OMS1}. However, there are the following two problems with this common approach.

1. The cavity modes need some time to evolve to their stable values. Hence, the transient dynamical process at the beginning cannot be described by the steady state expansion approach, while the final cooling limit just mainly depends on the whole evolution process \cite{cooling}. Thus, steady state expansion approach used in the past is not sufficient for exploring the cooling limit.

2. For the MIM system, the time-independent steady states only exist in some special cases, for example, $\Delta_1=\Delta_2=\omega_m$ or $\Delta_1=-\Delta_2=\omega_m$. The cavity mode oscillates with time in most cases and the steady state expansion approach is obviously not applicable there. 

\begin{figure*}[tbp]
\centering\includegraphics[width=15.5cm]{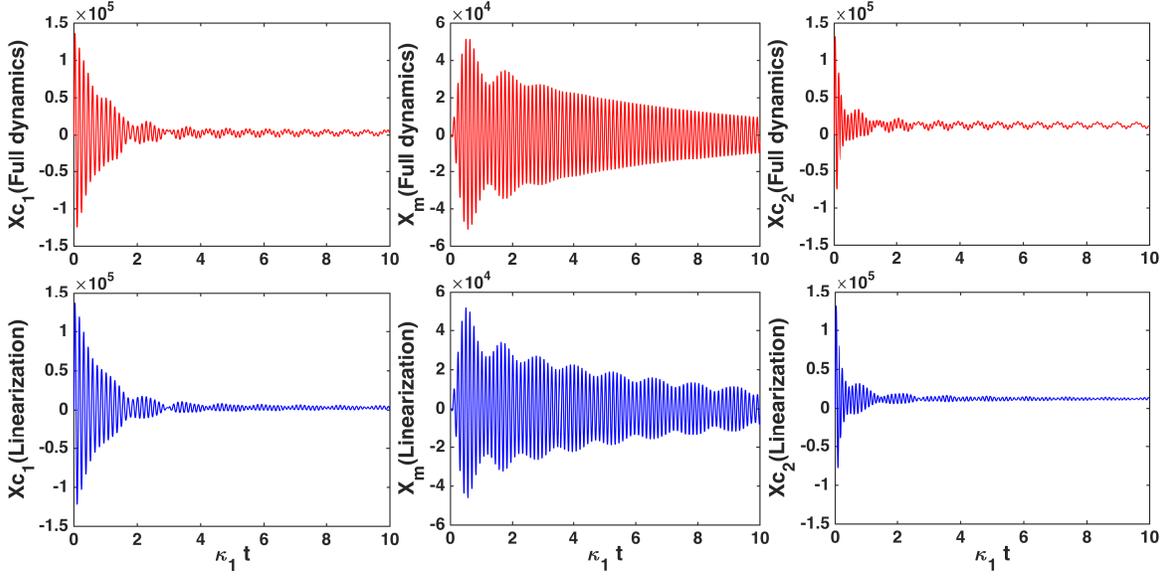}
\caption{Comparisons of the quadratures of two cavities $X_{c_1}(t)$, and $X_{c_2}(t)$, the displacement of mechanical resonator $X_m(t)$ predicted with the nonlinear dynamical equations Eq. (\ref{neq}) (red lines) and our linearized equations Eq. (\ref{dye1}) (blue lines). The parameters are $\kappa_2/\kappa_1=5$, $g_m/\kappa_1=10^{-5}$, $\omega_m/\kappa_1=50$, $\gamma_m/\kappa_1=10^{-3}$, $\Delta_1/\kappa_1=45$, $\Delta_2/\kappa_1=55$, $E_1/\kappa_1=4.5\times10^6$, $E_2/\kappa_1=5.5\times10^6$, and $J/\kappa_1=1$.}  \label{com}
\end{figure*}

To make the second point clearer, we use the above nonlinear quantum Langevin equations to predict the dynamical quantities of MIM system, such as the quadratures $X_{c_{1(2)}}=\frac{1}{\sqrt{2}}\left\langle\hat{a}_{1(2)}(t)+\hat{a}_{1(2)}^\dag(t)\right\rangle$ and the displacement $X_m=\frac{1}{\sqrt{2}}\left\langle\hat{b}(t)+\hat{b}^\dag(t)\right\rangle$, by assuming the factorization of the nonlinear factors, e.g., $\left\langle\hat{b}\hat{a}_1\right\rangle=\left\langle\hat{b}\right\rangle\left\langle\hat{a}_1\right\rangle$ \cite{eit}. The prediction of these quantities with the above nonlinear equations under the conditions, $\kappa_2/\kappa_1=5$, $\omega_m/\kappa_1=50$, $\Delta_1/\kappa_1=45$, $\Delta_2/\kappa_1=55$, and $J/\kappa_1=1$, are displayed in the first row of Fig. \ref{com}. With oscillation of the quadratures $X_{c_{1(2)}}$, the cavity modes will not become time-independently steady even after a long time, so it is not appropriate to use the linearization that is based on the steady state of the system.

Here, we adopt another approach developed in \cite{cooling} to solve the problem. Similar methods were also applied to some other physical systems \cite{He,Lin1,rydberg, Lin2,noise, dynamics, pulse1, Lin3, square}. First, we apply a decomposition as follows:
\begin{align}
U(t)=&\mathcal{T}\exp\{-i\int_0^t d\tau H_S(\tau)\}\nonumber\\
& \times\mathcal{T}\exp\{-i\int_0^t d\tau [H_{eff}(\tau)+H_N(\tau)]\},
\end{align} 
where $H_{eff}(\tau)+H_N(\tau)=\mathcal{T}\exp\{i\int_0^\tau dt' H_S(t')\}\{ H_{OM}+H_{SR}(\tau)\}\mathcal{T}\exp\{-i\int_0^\tau dt' H_S(t')\}$. 
The forms of the cavity and mechanical operators in the effective Hamiltonian $H_{eff}(\tau)$ and $H_N(\tau)$ are found as
\begin{widetext}
\begin{align}
&\mathcal{T}\exp\{i\int_0^t d\tau H_S(\tau)\}~\hat{a}_1~\mathcal{T}\exp\{-i\int_0^t d\tau H_S(\tau)\}= e^{-i\omega_{c_1} t}(\hat{a}_1+\frac{iE_1}{\Delta_1}(1-e^{i\Delta_1 t}))\equiv e^{-i\omega_{c_1} t}(\hat{a}_1+E_1(t)),\nonumber\label{trf}\\
&\mathcal{T}\exp\{i\int_0^t d\tau H_S(\tau)\}~\hat{a}_2~\mathcal{T}\exp\{-i\int_0^t d\tau H_S(\tau)\}= e^{-i\omega_{c_2} t}(\hat{a}_2+\frac{iE_2}{\Delta_2}(1-e^{i\Delta_2 t}))\equiv e^{-i\omega_{c_2} t}(\hat{a}_2+E_2(t)),\nonumber\\
&\mathcal{T}\exp\{i\int_0^t d\tau H_S(\tau)\}~\hat{b}~\mathcal{T}\exp\{-i\int_0^t d\tau H_S(\tau)\}=e^{-i\omega_mt}\hat{b}, 
\end{align}
where $E(t)$ is a time-dependent function. Therefore, we have
\begin{align}
&H_{eff}(t)=\nonumber\\
&-g_m\left[E_1(t)\hat{a}_1^{\dag}+E_1^*(t)\hat{a}_1+|E_1(t)|^2\right](e^{-i\omega_mt}\hat{b}+e^{i\omega_mt}\hat{b}^{\dag}) +g_m\left[E_2(t)\hat{a}_2^{\dag}+E_2^*(t)\hat{a}_2+|E_2(t)|^2\right](e^{-i\omega_mt}\hat{b}+e^{i\omega_mt}\hat{b}^{\dag}) \nonumber\\
&+\left[Je^{i(\omega_{c_2}-\omega_{c_1})t}(\hat{a}_2^{\dag}+E_2^*(t))(\hat{a}_1+E_1(t))+H.c.\right]+i\sqrt{2\kappa_1}\left\{e^{i\omega_{c_1} t}\left(\hat{a}_1^{\dag}+E_1^*(t)\right) \hat{\xi}_{c_1}(t)-H.c.\right\}\nonumber\\
&+i\sqrt{2\kappa_2}\left\{e^{i\omega_{c_2} t}\left(\hat{a}_2^{\dag}+E_2^*(t)\right) \hat{\xi}_{c_2}(t)-H.c.\right\}+i\sqrt{2\gamma_m}\left(e^{i\omega_mt}\hat{b}^{\dag}\hat{\xi}_m(t)-H.c.\right),\nonumber\\
&H_N(\tau)=-g_m(\hat{a}_1^{\dag}\hat{a}_1-\hat{a}_2^{\dag}\hat{a}_2)(e^{-i\omega_m\tau}\hat{b}+e^{i\omega_m\tau}\hat{b}^{\dag}).
\end{align} 
The effective Hamiltonian $H_{eff}(t)$ leads to the linearized equations of motion:
\begin{align}
\dot{\hat{a}}_1=&-\kappa_1 \hat{a}_1+i g_m E_1(t) (e^{-i\omega_mt}\hat{b}+e^{i\omega_mt}\hat{b}^{\dag})-iJe^{-i(\omega_{c_2}-\omega_{c_1})t}\hat{a}_2 \underbrace{-iJe^{-i(\omega_{c_2}-\omega_{c_1})t}E_2(t)-\kappa_1 E_1(t)}_{\lambda_1(t)}+\underbrace{\sqrt{2\kappa_1}e^{i\omega_{c_1} t}\hat{\xi}_{c_1} (t)}_{n_1(t)},\nonumber\\
\dot{\hat{b}}=&-\gamma_m \hat{b}+i g_m e^{i\omega_m t}\left[E_1(t)\hat{a}_1^{\dag}+E_1^*(t)\hat{a}_1-E_2(t)\hat{a}_2^{\dag}-E_2^*(t)\hat{a}_2\right] +\underbrace{i g_m e^{i\omega_m t}(|E_1(t)|^2-|E_2(t)|^2)}_{\lambda_2(t)}+\underbrace{\sqrt{2\gamma_m}e^{i\omega_mt}\hat{\xi}_m(t)}_{n_2(t)},\nonumber\\
\dot{\hat{a}}_2=&-\kappa_2 \hat{a}_2-i g_m E_2(t) (e^{-i\omega_mt}\hat{b}+e^{i\omega_mt}\hat{b}^{\dag})-iJe^{i(\omega_{c_2}-\omega_{c_2})t}\hat{a}_1 \underbrace{-iJe^{i(\omega_{c_2}-\omega_{c_2})t}E_1(t)-\kappa_2 E_2(t)}_{\lambda_3(t)}+\underbrace{\sqrt{2\kappa_2}e^{i\omega_{c_2} t}\hat{\xi}_{c_2} (t)}_{n_3(t)},\label{dye1}
\end{align}
while the effect of $H_N(\tau)$ is neglected under the condition $g_m/\omega_m \ll 1$ \cite{cooling}. The above equations can be rewritten as 
\begin{align}
\frac{d}{dt}\hat{\vec{c}}(t)=M(t)\hat{\vec{c}}(t)+\vec{\lambda}(t)+\hat{\vec{n}}(t),\label{dye}
\end{align}
in which we use the definitions $\hat{\vec{c}}=\{\hat{a}_1,\hat{a}^{\dag}_1,\hat{b},\hat{b}^{\dag},\hat{a}_2,\hat{a}^{\dag}_2\}^T$, $\vec{\lambda}(t)=\{\lambda_1(t),\lambda^*_1(t),\lambda_2(t),\lambda^*_2(t),\lambda_3(t),\lambda^*_3(t)\}^T$ and $\vec{n}(t)=\{n_1(t),n^*_1(t),n_2(t),n^*_2(t),n_3(t),n^*_3(t)\}^T$. The corresponding transformation matrix takes the form
\begin{align}
&M(t)=\begin{pmatrix}
 -\kappa_1& 0 & P_1(t) e^{-i\omega_m t}  & P_1(t) e^{i\omega_m t} &J(t) &0\\
 0& -\kappa_1 & P^*_1(t) e^{-i\omega_m t} & P^*_1(t) e^{i\omega_m t} &0 & J^*(t)\\
-P^{*}_1(t) e^{i\omega_m t} &P_1(t) e^{i\omega_m t}  & -\gamma_m & 0 &P_2^*(t) e^{i\omega_m t} &-P_2(t) e^{i\omega_m t}  \\
 P^*_1(t) e^{-i\omega_m t} & -P_1(t) e^{-i\omega_m t}  & 0  & -\gamma_m &-P_2^*(t) e^{-i\omega_m t}&P_2(t) e^{-i\omega_m t} \\
-J^*(t) & 0&-P_2(t)e^{-i\omega_m t}&-P_2(t)e^{i\omega_m t}&-\kappa_2 & 0\\
 0 & -J(t)&-P^*_2(t)e^{-i\omega_m t}&-P^*_2(t)e^{i\omega_m t}&0&-\kappa_2
\end{pmatrix},\label{tmatrix}
\end{align}
where $P_{1(2)}(t)=ig_mE_{1(2)}(t) $ and $J(t)=-iJe^{-i(\omega_{c_2}-\omega_{c_1})t}$. The solution of the dynamical equation, Eq. (\ref{dye}), is found as 
\begin{align}
\hat{\vec{c}}(t)&=\mathcal{T}\exp\{\int_0^t d\tau M(\tau)\}\hat{\vec{c}}(0)+\int_0^t d\tau \mathcal{T}\exp\{\int_\tau^t dt' M(t')\}(\vec{\lambda}(\tau)+\hat{\vec{n}}(\tau))\nonumber\\
&\equiv\hat{\vec{c}}_s(t)+\vec{c}_{ds}(t)+\hat{\vec{c}}_n(t), \label{solution}
\end{align}
in which we define
\begin{align}
&\mathcal{T}\exp\{\int_\tau^t dt' M(t')\}=\begin{pmatrix}
 d_{11}(t,\tau)& d_{12}(t,\tau) & d_{13}(t,\tau) & d_{14}(t,\tau)&d_{15}(t,\tau)&d_{16}(t,\tau) \\
 d_{21}(t,\tau)& d_{22}(t,\tau) & d_{23}(t,\tau) & d_{24}(t,\tau)& d_{25}(t,\tau)& d_{26}(t,\tau) \\
d_{31}(t,\tau)& d_{32}(t,\tau) & d_{33}(t,\tau) & d_{34}(t,\tau)& d_{35}(t,\tau)& d_{36}(t,\tau) \\
 d_{41}(t,\tau)& d_{42}(t,\tau) & d_{43}(t,\tau) & d_{44}(t,\tau)& d_{45}(t,\tau)& d_{46}(t,\tau)\\
  d_{51}(t,\tau)& d_{52}(t,\tau) & d_{53}(t,\tau) & d_{54}(t,\tau)& d_{55}(t,\tau)& d_{56}(t,\tau)\\
   d_{61}(t,\tau)& d_{62}(t,\tau) & d_{63}(t,\tau) & d_{64}(t,\tau)& d_{65}(t,\tau)& d_{66}(t,\tau)\\
\end{pmatrix}.
\end{align}
\end{widetext}
The three terms in Eq. (\ref{solution}) are respectively the contributions from the initial values, the coherent drives, and the noise drives.
Now, with the numerical solution of the above equations, one can obtain the real-time evolution of system quantities such as the quadratures, displacement, evolved photon number, thermal phonon number, etc. 

Using the approach detailed in the above, we calculate the evolved cavity quadratures $X_{c_{1}}$, $X_{c_{2}}$ and the displacement $X_m$ of mechanical resonator. The predictions of these quantities with our linearization approach match well with those predicted by the nonlinear Langevin equations, as shown in Fig. (\ref{com}). In particular, the oscillation feature of the system quantities can be well captured by our approach. Such match implies that there would be no big difference of other quantities, including the evolved thermal phonon number, between those predicted by our approach 
and the real values. In this sense, our linearization approach can be well applied to the concerned dynamical processes and find the possible limit for cooling.

\section{cooling under two red-detuned drive fields}
\label{sec3}
\begin{figure*}[t!]
\centering\includegraphics[width=13.5cm]{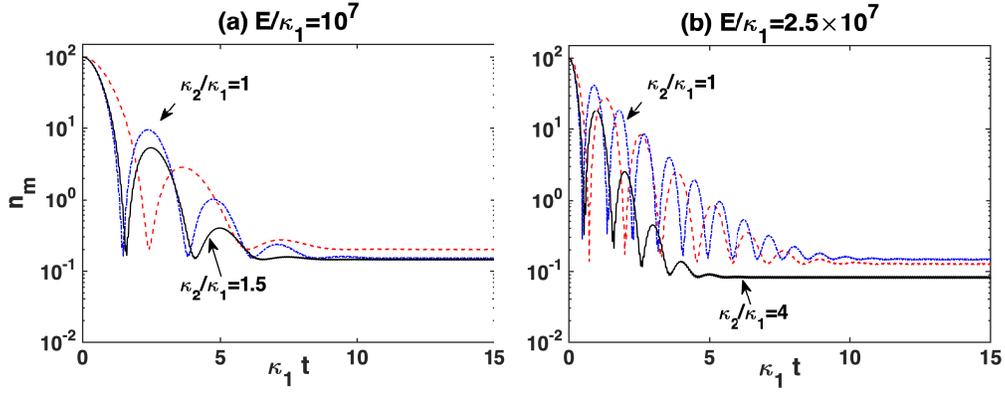}
\caption{Evolution of thermal phonon number in a MIM system under different drive intensities $E/\kappa_1$. The evolution with the corresponding single-cavity setup (denoted by red dashed line) is presented for comparison.  In (a) the final phonon 
number with the same cavity damping rate as in the MIM system, i.e., $\kappa_1=\kappa_2$ (blue dash-dotted line), is lower than the one achieved with the single-cavity setup, while in (b) it is higher than the one with single cavity. With the proper choice of the second cavity damping rate, e.g., $\kappa_2/\kappa_1=1.5$ in (a) and $\kappa_2/\kappa_1=4$ in (b) (black solid line), the final phonon number can be always lower than the one of single-cavity system. The other parameters are chosen as $g_m/\kappa_1=10^{-5}$, $\Delta_1/\kappa_1=\Delta_2/\kappa_1=\omega_m/\kappa_1=100$, $\gamma_m/\kappa_1=10^{-3}$, and $n_{th}=100$.}  \label{evo1}
\end{figure*}
Now we discuss the cooling effect of MIM system. A director indicator of the cooling effect is the evolved thermal phonon number, $n_m(t)=\langle\hat{b}^{\dag}\hat{b}(t)\rangle-\langle\hat{b}^{\dag}(t)\rangle\langle\hat{b}(t)\rangle$, which can be found directly with the numerical solution of Eq. (\ref{solution}). The pure drive terms $\vec{c}_{ds}(t)$ contributes to $\langle\hat{b}(t)\rangle$, which exhibits as the kinetic energy of a mechanical resonator, but it does not change the thermal property of the resonator, 
so its contribution $\langle\hat{b}^{\dag}(t)\rangle\langle\hat{b}(t)\rangle$ should be excluded in the calculation of the thermal phonon number $n_m(t)$. 
Therefore, the quantity $n_m(t)$ is determined by two parts; one is the contribution 
\begin{align}
&\langle \hat{b}^{\dag}(t)\hat{b}(t)\rangle_s  \nonumber\\
&=d_{32}(t,0)d_{41}(t,0)+d_{33}(t,0)d_{44}(t,0)n_{th}\nonumber\\
&+d_{34}(t,0)d_{43}(t,0)(n_{th}+1)+d_{36}(t,0)d_{45}(t,0)
\end{align}
from the initial condition of the system, and the other is the contribution 
\begin{align}
&\langle\hat{b}^{\dag}(t) \hat{b}(t)\rangle_r  \nonumber\\
&=\int_0^t d\tau \left[2\kappa_1d_{32}(t,\tau)d_{41}(t,\tau)+2\gamma_m n_{th} d_{33}(t,\tau)d_{44}(t,\tau)\right. \nonumber\\
&\left.+2\gamma_m(n_{th}+1) d_{34}(t,\tau)d_{43}(t,\tau)+2\kappa_2 d_{36}(t,\tau)d_{45}(t,\tau)\right]
\end{align}
from the noises that satisfy the relations $\langle\hat{\xi}_c(t)\hat{\xi}^{\dag}_c(t')\rangle=\delta(t-t'),
\langle\hat{\xi}^{\dag}_c(t)\hat{\xi}_c(t')\rangle=0,
\langle\hat{\xi}_m(t)\hat{\xi}^{\dag}_m(t')\rangle=(n_{th}+1)\delta(t-t'),
\langle\hat{\xi}^{\dag}_m(t)\hat{\xi}_m(t')\rangle=n_{th}\delta(t-t')$.

Because our aim is to cool down the mechanical resonator, the detuning of two drives will be set to be the same as $\Delta_1=\Delta_2=\omega_m$, which is the resonance point to achieve the maximum beam splitter (BS) effect. Without loss of generality, we here set the amplitude of two drives to be the same as $E_1=E_2=E$ and the tunneling $J=0$ (the relation between the cooling effect and tunneling will be discussed in next section). The evolution of phonon number from the initial one $n_{m}=100$ with $E/\kappa_1=10^7$ or $2.5\times10^7$ is displayed in Fig. \ref{evo1}. The red dashed line represents the results for single cavity, which is presented for comparison. Two cases with different combinations of cavity damping are denoted by blue and black lines, respectively. The mechanical resonator can be efficiently cooled down to the ground state with the conditions in Fig. \ref{evo1}. 

Fig. \ref{evo1}(a) specifically shows that, with a MIM system, the phonon number is rapidly lowered to that of a ground state and the corresponding stable phonon number is lower than the one that can be achieved with a single-cavity system. This implies that the cooling effect can be efficiently improved by the MIM system. In addition, the character that higher cooling speed corresponds to stronger cooling effect agrees well with that in Ref. \cite{cooling}. On the other hand, as seen from Fig. \ref{evo1}(b), a lower phonon number can be achieved under the conditions $\kappa_2/\kappa_1=4$ and $E/\kappa_1=2.5\times10^7$ in the MIM system. However, cooling can not be always improved by MIM system, for example, in the case of the same cavity damping rate $\kappa_1=\kappa_2$. A suitable choice of the cavity damping rates $\kappa_2$ and the drive intensity $E/\kappa_1$ is therefore crucial to improving the cooling effect, as we will discuss in the following section.

\section{cooling effect dependence on system parameters}
\label{sec4}

\begin{figure*}[btp]
\centering\includegraphics[width=11.5cm]{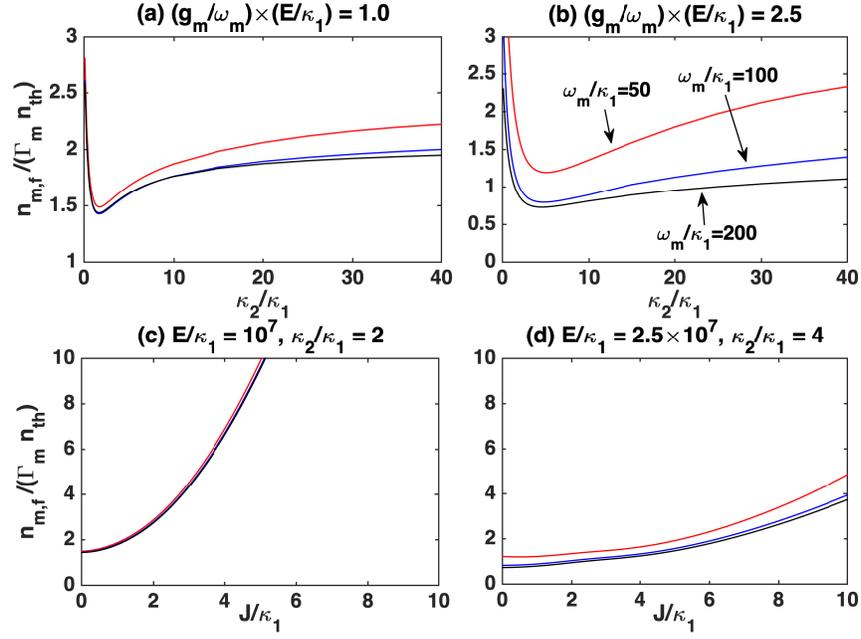}
\caption{Relations between the cooling ratio $n_{m,f}/\Gamma_m n_{th}$ and the second cavity damping rate $\kappa_2/\kappa_1$ or the tunneling strength $J/\kappa_1$. In each figure, three cases with different mechanical frequencies $\omega_m/\kappa_1=50$, $100$, and $200$ are displayed. From (a) and (b), we find that for the fixed effective drive intensity $(g_m/\omega_m)\times(E/\kappa_1)$, there is an optimum damping rate, $\kappa_2/\kappa_1$, for which the cooling ratio is the smallest. From (c) and (d), we conclude that tunneling will not help improve the cooling effect. The other parameters are chosen as $g_m/\kappa_1=10^{-5}$, $\Delta_1/\kappa_1=\Delta_2/\kappa_1=\omega_m/\kappa$, $\gamma_m/\kappa_1=10^{-3}$, and $n_{th}=100$.}  \label{cp}
\end{figure*}

Next we discuss how the system conditions, including the cavity damping ratio $\kappa_2/\kappa_1$, the tunneling rate $J$, and the drive intensity $E$, affect the cooling effect. The cooling ratio $n_{m,f}/(\Gamma_m n_{th})$ with $\Gamma_m=\gamma_m/\kappa_1$ is used here as the figure of merit for the cooling effect. The first parameter we consider is the cavity damping rate. The cavity damping rate  $\kappa_1$ of the left cavity is set to be fixed. The relations between the cooling effect and the cavity damping ratio $\kappa_2/\kappa_1$ are presented in Fig. \ref{cp}(a) and \ref{cp}(b), with the different effective drive intensity values $(g_m/\omega_m)\times(E/\kappa_1)=1, 2.5$. When the cavity damping ratio is small, e.g., $\kappa_2/\kappa_1=2$ as in Fig. \ref{cp}(a), the cooling ratio will be high. It means that the small damping rate $\kappa_2/\kappa_1$ does not help the cooling, since the second drive is hard to be injected into the right cavity, resulting in a weak interaction (cf. the term $-\kappa_2E_2(t)$ in Eq. (\ref{dye1})). On the other hand, when the cavity damping ratio is high, e.g., $\kappa_2/\kappa_1=100$, the cooling ratio will also become high. If $\kappa_2/\kappa_1\rightarrow \infty$, the second drive field will appear to be non-existing, and the cooling ratio will therefore tend to be the same as the one for only single cavity under one drive \cite{cooling}. Moreover, each of the curves exhibits a dip indicating that there is an optimum choice of cavity damping rate $\kappa_2/\kappa_1$ for the given effective drive intensity $(g_m/\omega_m)\times(E/\kappa_1)$. The existence of the dip reflects the competition between the cavity damping, which is not beneficial to cooling, and the interaction between the drive fields and mechanical resonator, which benefits cooling. That is why a suitable combination of cavity damping ratio $\kappa_2/\kappa_1$ and the effective drive intensity $(g_m/\omega_m)\times(E/\kappa_1)$ is crucial to the cooling of mechanical resonator; see Fig. \ref{cp}.
\begin{figure}[btp]
\centering\includegraphics[width=8.5cm]{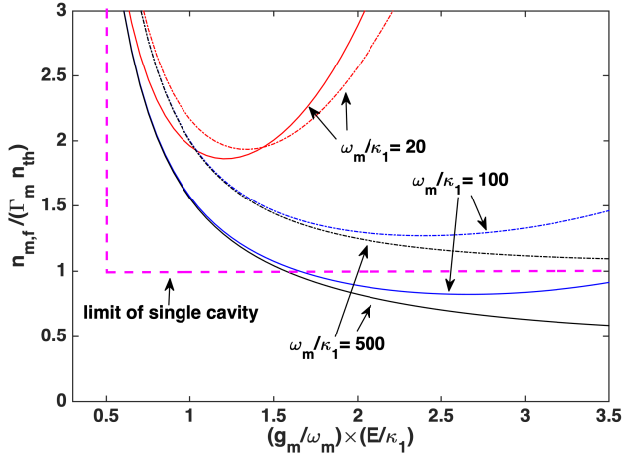}
\caption{Relation between the cooling ratio $n_{m,f}/(\Gamma_m n_{th})$ and the effective drive intensity $(g_m/\omega_m)\times(E/\kappa_1)$ with different mechanical frequency $\omega_m/\kappa_1=20$, $100$, $500$. The dash-dotted lines represent the relations in case of single cavity; which are used for comparison. The pink dashed line is the cooling limit of the single cavity with $\omega_m/\kappa_1\rightarrow \infty$. It is obvious that the cooling effect can be efficiently improved by a coupled cavity system. In particular, the improvement is significant with large mechanical resonator such as the one with $\omega_m/\kappa_1=100$ or $500$, and the cooling ratio with it can be even lower than the cooling limit of the single cavity. The system parameters are set as $g_m/\kappa_1=10^{-5}$, $\Delta_1/\kappa_1=\Delta_2/\kappa_1=\omega_m/\kappa$, $\gamma_m/\kappa_1=10^{-3}$, and $\kappa_2/\kappa_1=4$.}  \label{cp2}
\end{figure}

The second parameter is the tunneling rate $J$. According to the relations in Figs. \ref{cp}(c) and \ref{cp}(d), the cooling ratio simply increases with increasing tunneling rate. It increases quickly when the drive intensity is relatively weak as shown in Fig. \ref{cp}(c). Obvious one cannot cool down the mechanical resonator efficiently if the tunneling of the two cavities is high. The reason is that tunneling actually lowers the interaction between the cavity modes and the mechanical oscillator. In order to achieve the expected cooling effect, tunneling should be rendered as weak as possible.

A more important parameter that affects the cooling effect is the drive intensity $E_2=E_1=E$. The factor $g_mE_{1(2)}(t)$ in Eqs. (\ref{dye1}) represents the effective intensity of two drives, so we use the dimensionless effective intensity $(g_m/\omega_m)\times(E/\kappa_1)$ in the discussion. The relations between the cooling ratio and the effective intensity are displayed in Fig. \ref{cp2}. The dashed lines represent the relations for a single-cavity setup used for comparisons. When the mechanical frequency is small, e.g., $\omega_m/\kappa_1=20$, the improvement by the MIM system is not significant, and the cooling ratio will increase when the effective intensity is higher than a certain value. This is because both squeezing (SQ) effect [indicated by the terms proportional to $\hat{a}^\dagger_{1(2)}$ and $\hat{b}^\dagger$ on the right sides of Eq. (\ref{dye1})] and BS effect will be enhanced with increased drive intensity. The competition between SQ effect and BS effect will finally decide the cooling ratio. When the mechanical frequency is relatively high, e.g., $\omega_m/\kappa_1=100$, the improvement will become significant and the cooling ratio can be lower than $n_{m,f}/(\Gamma_m n_{th})=1$, the theoretical limit of a single-cavity setup under the condition $\omega_m/\kappa\rightarrow \infty$ (red dashed line) \cite{cooling}. This feature is more obvious with a higher mechanical frequency, e.g., $\omega_m/\kappa_1=500$. The cooling ratio can approach the value of $0.5$ with the present parameters, 
and it can be further reduced with a higher drive intensity.

\section{Optimal cooling with MIM}
\label{sec5}

\begin{figure*}[btp]
\centering\includegraphics[width=15cm]{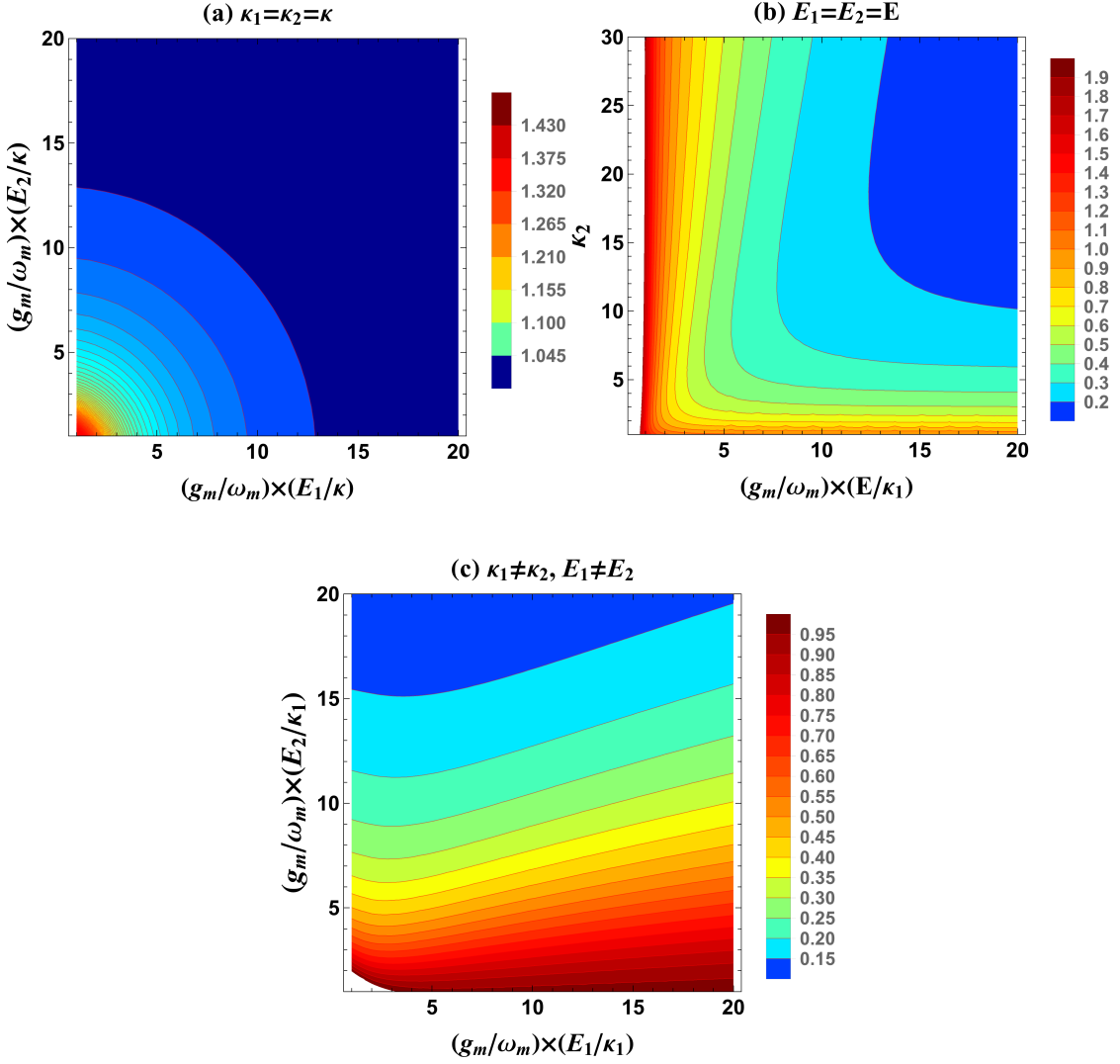}
\caption{Cooling ratio $n_{m,f}/(\Gamma_m n_{th})$ under different conditions. (a) $\kappa_1=\kappa_2=\kappa$. The circle boundary lines indicate that the cooling ratio mainly depends on the quadratic sum of the two effective drive intensities $J_{E_{1(2)}}=(g_m/\omega_m)\times(E_{1(2)}/\kappa)$. The cooling ratio approaches the limit $1$, which is the same as that with single cavity. (b) $\kappa_2\neq \kappa_1$, $E_1=E_2=E$. The cooling limit can be very small with the high effective drive intensity associated with optimum cavity damping rate $\kappa_2/\kappa_1$. (c) $\kappa_2\neq \kappa_1$, $E_1\neq E_2$. With the appreciated choice of the second cavity damping rate $\kappa_2/\kappa_1$, the cooling rate can be significantly reduced by increasing the second drive intensity. However, increasing the first drive intensity will not help improve the cooling effect at all; in contrast, it may worsen the scenario. The parameters, which are not indicated in the figures, are chosen as $g_m/\kappa_1=10^{-5}$, $\Delta_1/\kappa_1=\Delta_2/\kappa_1=\omega_m/\kappa_1$, and $\gamma_m/\kappa_1=10^{-3}$.}  \label{limit}
\end{figure*}

To find the cooling limit of the MIM system, we let the mechanical frequency $\omega_m/\kappa_1\rightarrow \infty$. In this limit, the elements in the transformation matrix $M(t)$ of Eq. (\ref{tmatrix}) take the following form.
\begin{align}
P_{1(2)}(t)e^{-i\omega_mt}&\rightarrow g_mE_{1(2)}/\omega_m,\nonumber\\
P_{1(2)}(t)e^{i\omega_mt}&\rightarrow 0.
\end{align}
With the condition $J=0$, the transformation matrix is now time-independent as $M$, leading to the reduction of the time-order exponential $\mathcal{T}\exp\{\int_\tau^t dt' M(t')\}$ to the normal exponential $\exp\{\int_\tau^t Mdt'\}$. The analytical solution of Eq. (\ref{solution}) can now be found as we will discuss three cases as follows, and the cooling ratio can be seen more clearly to be even lower by one order than the optimal result by using single cavity.

\subsection{$\kappa_2=\kappa_1$}
The first case is the cooling limit with the same cavity damping rate, i.e., $\kappa_2=\kappa_1$. Under this condition, we plot the relations between the cooling ratio $n_{m,f}/(\Gamma_m n_{th})$ and the two effective drive intensities $J_{E_{1(2)}}=(g_m/\omega_m)\times(E_{1(2)}/\kappa_1)$ as shown in Fig. \ref{limit}(a). The boundary lines are circular, which means that the cooling ratio depends on the quadratic sum of the two effective drive intensities, i.e., $J^2_{E_1}+J^2_{E_2}$. With the increase in drive intensity, the cooling ratio approaches $1$, which is the cooling limit with a single cavity. The exact same as the following analytical cooling limit can be obtained,
\begin{align}
\frac{n_{m,f}}{\Gamma_m n_{th}}= \frac{1+\Gamma_m+J^2_{E_1}+J^2_{E_2}}{(1+\Gamma_m)(\Gamma_m+J^2_{E_1}+J^2_{E_2})}.
\end{align}
It is obvious that when $J^2_{E_1}+J^2_{E_2}\gg 1$ and $\Gamma_m \ll 1$, the cooling limit $n_{m,f}/(\Gamma_m n_{th})\simeq1$. That does mean that the cooling effect cannot be effectively improved with the trivial combination of two same cavities, which has been mentioned in Sec. \ref{sec4} as well. 

\subsection{$\kappa_2\neq \kappa_1$, $E_1=E_2=E$}
The second case is the cooling limit with different cavity damping rates, i.e., $\kappa_2 \neq \kappa_1$. Here the two drive intensities are set to be the same $E_1=E_2=E$. Under these conditions, we plot the relations between the cooling ratio $n_{m,f}/(\Gamma_m n_{th})$ and the second cavity damping rate $\kappa_2$, together with the effective intensity $J_E=(g_m/\omega_m)\times(E/\kappa_1)$ as shown in Fig. \ref{limit}(b). Clearly, the cooling effect can be significantly improved with the appreciated choice of cavity damping rate and drive intensity. Even a very small cooling ratio, e.g., $n_{m,f}/(\Gamma_m n_{th})=0.1$, corresponding to the improvement of cooling effect by one order, is possible with the MIM system. Furthermore, when the effective drive intensity is as high as $J^2_E\gg1$, we find that the best choice of cavity damping rate $\kappa_2/\kappa_1$ is, 
\begin{align}
\kappa_2/\kappa_1 &= (1+\Gamma_m+\sqrt{24 J^2_E -3+ 2 \Gamma_m - \Gamma^2_m})/2 \nonumber\\
&\simeq (2\sqrt{6}J_E+1)/2,
\end{align}
with $\Gamma_m\ll1$. The corresponding cooling limit can be achieved as follows.
\begin{align}
\frac{n_{m,f}}{\Gamma_m n_{th}}\simeq\frac{4\sqrt{6}}{3J_E}\sim J^{-1}_E.
\end{align}
Compared to the cooling with single cavity, with a cooling limit of $n_{m,f}/(\Gamma_m n_{th})=1$, a very small cooling ratio can be obtained with coupled cavity system by increasing the effective intensity to a maximum possible value.

\subsection{$\kappa_2\neq \kappa_1$, $E_1\neq E_2$}
Finally, we discuss the general case of different cavity damping rates and different drive intensities as, $\kappa_2\neq \kappa_1=1$, $E_1\neq E_2$. Now, the cooling ratio will change with three parameters, $\kappa_2/\kappa_1$, $J_{E_1}$, and $J_{E_2}$. Through calculations, we find that the best choice of cavity damping rate $\kappa_2/\kappa_1$ takes the following form.
\begin{align}
\kappa_2/\kappa_1 &= (1+\Gamma_m+\sqrt{12 J^2_{E_1}+12 J^2_{E_2} -3+ 2 \Gamma_m - \Gamma^2_m})/2 \nonumber\\
&\simeq (1+2\sqrt{3 J^2_{E_1}+3 J^2_{E_2} })/2,
\end{align}
when the quadratic sum of two effective drive intensities is large as in $J^2_{E_1}+J^2_{E_2}\gg 1$ and $\Gamma_m\ll1$. Under this condition, we also plot the relations between the cooling ratio $n_{m,f}/(\Gamma_m n_{th})$ and the two effective drive intensities $J_{E_{1(2)}}=(g_m/\omega_m)\times(E_{1(2)}/\kappa_1)$ as shown in Fig. \ref{limit}(c). Obviously, the two drive fields will have different effect in improving the cooling effect. When the effective intensity $J_{E_1}$ of first drive field in the left cavity is fixed, the cooling ratio can be well reduced to a small value as expected, e.g., $n_{m,f}/(\Gamma_m n_{th})=0.1$, by increasing the second effect drive intensity $J_{E_2}$. However, when the effective intensity $J_{E_2}$ of second drive field in the right cavity is fixed, the cooling ratio cannot be reduced but it shows an increase with increasing first effective drive intensity $J_{E_1}$. This feature can be found clearly in the following approximate solution.
\begin{align}
\frac{n_{m,f}}{\Gamma_m n_{th}}\simeq \frac{4\sqrt{3}\sqrt{J^2_{E_1}+J^2_{E_2}}}{3J^2_{E_2}},
\end{align}
under the conditions $J^2_{E_1}+J^2_{E_2}\gg 1$ and $\Gamma_m\ll1$. The denominator only involves the second effective drive intensity $J_{E_2}$. Hence, the second drive intensity needs to be increased to the highest possible value to improve the cooling effect.

\section{Conclusion}
\label{conclusion}
In this work, we present a detailed study on cooling of a mechanical resonator with MIM system. Our approach applies to the general situation of such system, and is capable 
of predicting the evolutions of all concerned system quantities. We find that the displacement of resonator in the system will not generally become time-independently stable in the end, so that the dynamical behavior of such systems cannot be well explained with the steady state approach adopted in the past \cite{OMS1}. The cooling processes with such systems are more appropriately explored in a dynamical approach applied in the current work. Using the approach, one is able to see the detailed cooling processes under various conditions determined by the system parameters (such as damping rates, tunneling rate, and drive intensity) and prepare the systems properly for achieving a good cooling result that breaks the cooling limit by the single-cavity (single-drive) setups. Our numerical calculations indicate that the improvement of cooling by one order of the minimum thermal phonon number is possible, as compared with the systems of single cavity. The use of the MIM systems with high sideband resolution $\omega_m/\kappa_1$ may 
provide a path to ultra-efficient cooling of mechanical resonator.

\section*{Acknowledgment}

This project is sponsored by National Natural Science Foundation of China (NSFC) (11574093); Natural Science Foundation of Fujian Province of China (NSFFPC) (2017J01004); Promotion Program for Young and Middle-aged Teacher in Science and Technology Research of Huaqiao University (PPYMTSTRHU) (ZQN-PY113).

\end{document}